# Hydroxylation of Rutile TiO$_2$(110) Surface Enhancing Its Reducing Power for Photocatalysis

*Daoyu Zhang,*[†] *Minnan Yang*[‡]*, Shuai Dong\**[,†]

[†]Department of Physics, Southeast University, Nanjing, 211189, China.

[‡]Department of Physics, China Pharmaceutical University, Nanjing, 211198, China.

**ABSTRACT:** Hydroxylation of the rutile TiO$_2$(110) surface has attracted much attention as the excess unpaired electrons introduced by hydroxyls play a critical role in surface chemistry and photocatalysis process of this material. In this work, based on density functional theory calculations with the Hubbard U correction, the electronic structures of the hydroxylated TiO$_2$(110) surfaces have been studied. One interesting effect is found that the hydroxylation can elevate band edges of TiO$_2$, and thus can enhance its reducing power for photocatalysis. The underlying physical mechanism for such shifts of the band edges are associated with the electric dipoles arising from the hydroxyl groups on the surface.

**INTRODUCTION**

Many so-called 'clean' single-crystalline oxide surfaces, obtained with usual cleaning procedures, are more or less covered by hydrogen atoms in the form of hydroxyls, because water readily dissociates on oxygen vacancies to form pairs of nearby bridging hydroxyl groups.[1-5] As an important technological material, $TiO_2$ has many applications, such as photochemical hydrogen production from water, medical implants, and waste water treatment. In all these situations, $TiO_2$ surface contacts with water, therefore the presence of hydroxyls at the interface between $TiO_2$ and water is inevitable, which is one of the most common point defects on the $TiO_2$ surfaces. The high hydroxyl coverage on $TiO_2$ surfaces can be reached by hydrogenation in the $H_2$ or atomic H atmosphere at high temperature or even at room temperature.[6-8] Both experimental and theoretical works have demonstrated that atomic hydrogen readily sticks to the bridging oxygen of $TiO_2$ surface forming hydroxyl.[9-12] An interesting effect of hydrogenation is that the hydrogenated $TiO_2$ and other metal oxides usually show higher photoactivity and solar-to-hydrogen conversion efficiency.[13-18] In the past years, several theoretical studies have proposed a few mechanisms to explain the enhanced solar-to-hydrogen conversion efficiency of hydrogenated $TiO_2$, such as surface distortions created by added H atoms,[19] different natures between different hydrogenated surfaces,[11] and localized mid-gap states within the forbidden band.[20] Even though, the widely-accepted scenario has not been established yet.

The rutile $TiO_2$(110) surface is the thermodynamically most stable crystal face and therefore represents the dominating facet of rutile crystallities.[21] Hydroxyls on the $TiO_2$(110) surface give rise to excess electrons, which are localized on Ti atoms nearby and reduce their valence from a nominal +4 to +3. The unpaired excess electron trapped by a Ti atom induces the local distortion around due to strong Coulomb and exchange interactions, described as a polaronic structure. There have been a number of studies focusing on the characterization of polarons[22-26] and the effects of polarons on the surface chemistry and photochemistry of $TiO_2$.[4, 27-31] For example, the polarons create the mid-gap states in the electronic structure of the hydroxylated $TiO_2$ surface,[32] which can effectively enhance visible light absorption of $TiO_2$.[33] However, other effects of hydroxylation to the electronic structure of $TiO_2$, such as modulations of the band gap and the band edges, have not been studied before. It is well known that, to use a semiconductor and drive reactions with light, the band gap and band edges of this semiconductor are essential properties.[34-36] For example, to decompose water ($H_2O \rightarrow H_2$ + $1/2O_2$), the band gap of a semiconductor used as the photocatalyst must be greater than 1.23 eV, and the valence and conduction band edge energies must straddle the electrochemical potentials for two redox half reactions of water decomposition (see Figure 1). The reducing and oxidizing powers of the semiconductor for a half reaction are measured by the conduction and valence band edge energies respectively: the higher the conduction band edge energy and the lower valence band edge energy, the stronger tendency for reduction and oxidation.

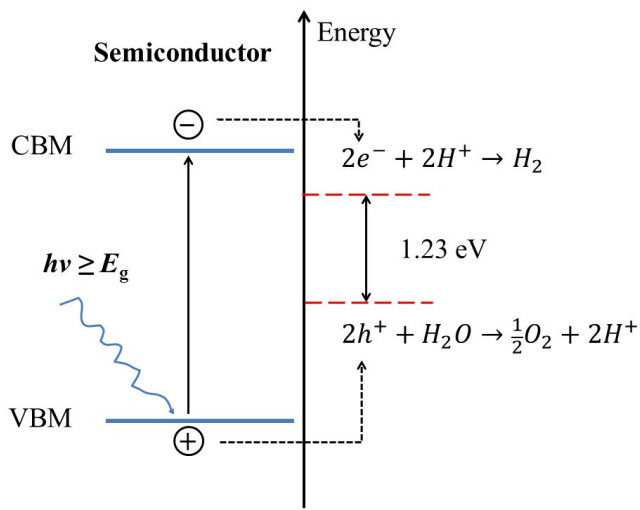

Figure 1. Illustration of a semiconductor driving the water decomposition reaction under illumination. The band edges of the semiconductor must straddle potentials for the hydrogen and oxygen evolution reactions.

Rutile $TiO_2$ is a semiconductor with a wide band gap of 3.0 eV, greater than the practical need for 1.6-2.4 eV to effectively drive water splitting.[36] Therefore, to use the solar radiation, many efforts have been aimed at the band gap narrowing of $TiO_2$.[34, 35, 37] Besides, not only the band gap, but also the band edges are crucial in redox reactions. In fact, not both of two band edge positions of rutile $TiO_2$ are suitable to drive two redox half reactions of water decomposition.[38, 39] From the thermodynamic consideration, there is no problem for holes of the valence band to drive the oxygen evolution reaction since the hole energy is lower than the electrochemical potential for this oxidation reaction.

However, its conduction band edge is only slightly greater than the potential of the $H^+/H_2$ redox couple, implying a weak reducing power of the conduction band electrons to produce the hydrogen gas.

In this work, using the density functional theory with Hubbard U correction (DFT+U) calculations, we demonstrate that hydroxylation of $TiO_2$ can lift both the conduction/valence band edges in a synchronous manner, suggesting that the reducing power increases and the hydrogen evolution reaction is promoted to be more efficient. The blue shifts of the band edges is electrostatic in origin and results from the dipoles created by the presence of hydroxyls at the $TiO_2$ surface. Our calculation may provide a reasonable explanation for higher solar-to-hydrogen conversion efficiency of the hydrogenated $TiO_2$.

**METHODOLOGY**

The band structure calculations were performed using the projector-augmented wave pseudopotentials as implemented in the Vienna ab initio Simulation Package (VASP)[40, 41]. The electronic interactions are described within the GGA+U formalism, where the Hubbard-type correction was applied on Ti's 3d orbitals. U = 3.5 eV for Ti's 3d orbitals was chosen in the present work based on the previous works that suggested that this U value could give a correct description of the polaronic states of the reduced $TiO_2$ and the interaction between hydrogen and the $TiO_2$ surfaces.[33] The energy cutoff for plane wave basis was set to be 450 eV. The atomic positions and cell parameters were relaxed until the forces on each atom are less than 0.01 eV/Å, and the self-consistent convergence accuracy was set at $1\times10^{-5}$ eV.

The (110) surface of $TiO_2$ was modeled by (3×2) periodically repeated slabs of four trilayers, in which the atoms of bottom two layers were fixed at their bulk positions[10] (see Figure 2a), leaving the top two layers to be the reconstructive surface layers. To simulate the open surface, there is a vacuum space between the slab and its periodic replicas along the [001] direction. At the surface, the coordinations of O and Ti are reduced. As shown in Figure 2a and 2b, the rows of twofold-coordinated bridging O atoms ($O_{2c}$) are parallel to the rows of fivefold-coordinated Ti atom along the [100] direction. Hydrogen adsorption on the surface was modelled by adding neutral H atoms to the $O_{2c}$ atoms. While Γ-point sampling was used for geometrical relaxation of surfaces, automatically generated Γ-centered 3×2×1 Monkhorst-Pack mesh was used for all electronic structure calculations. The monopole, dipole and quadrupole corrections have been applied to the electrostatic interaction between the slab and its periodic images in the direction perpendicular to the slab.

To compare the Kohn-Sham levels from different DFT calculations, we select the vacuum level as the common zero energy reference (shifted to 0 eV) for the bands energy alignment.[42-44] The vacuum level of a surface is determined by the (110)-planar average electrostatic potentials in the vacuum region, where the potentials (denoted by $E_v$) are not varied as the [001] direction. Then the eigenenergy of every state subtracting $E_v$ is the level of this state.

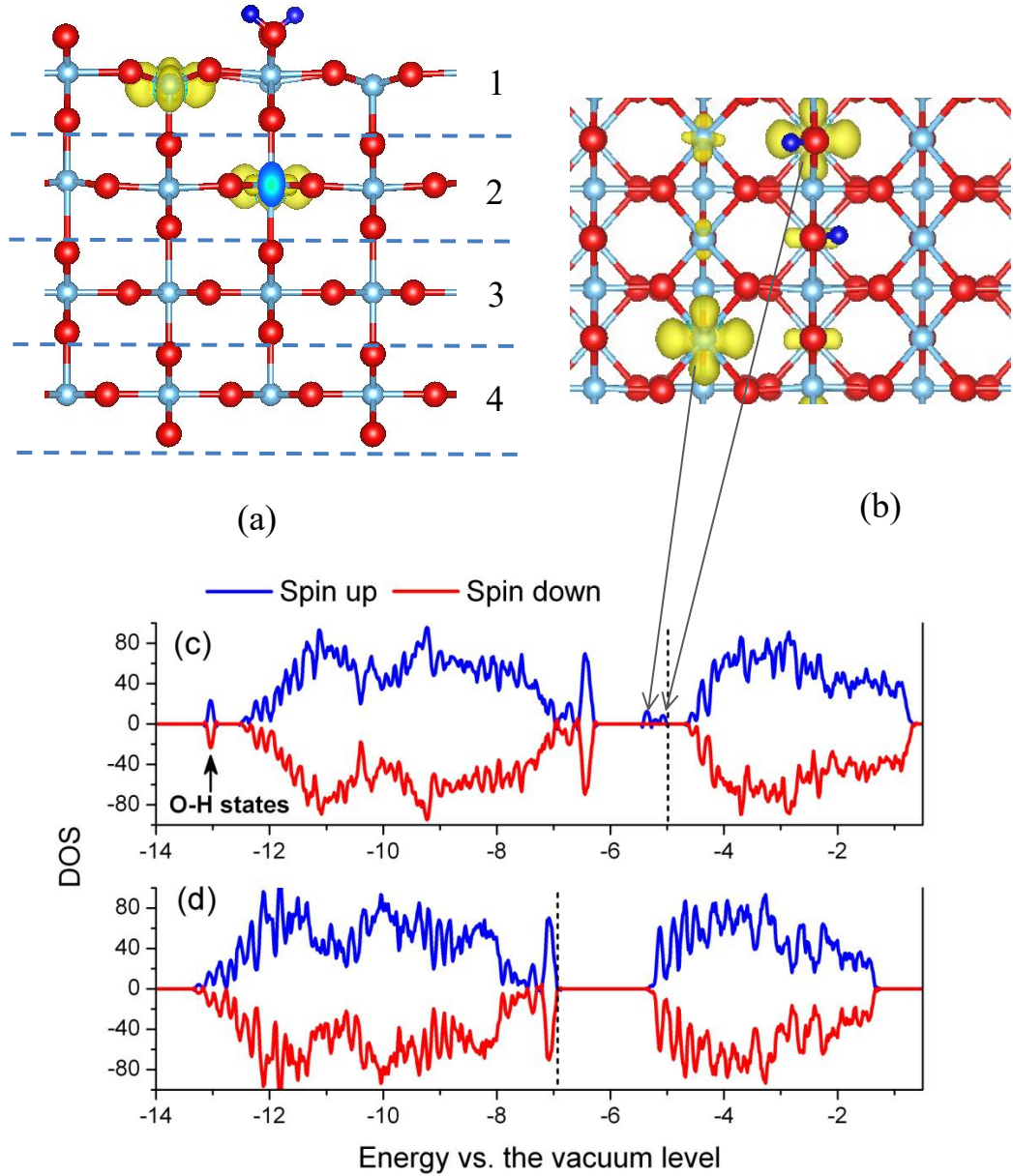

Figure 2. (a) The side view and (b) the top view of the rutile $TiO_2(110)$ surface with two nearby hydroxyls, in which the spin charge density is plotted (golden lobes). Cyan: Ti;

red: O; blue: H. The DOS of the (110) surface (c) with and (d) without two nearby hydroxyls respectively, where the vertical, short dash lines represent the Fermi level.

## RESULTS AND DISCUSSION

**Polaronic states of the hydroxylated surface.** For comparison with the results from previous studies, we first present the defect states of the rutile $TiO_2(110)$ surface with two nearby hydroxyls as shown in Figure 2. The spin charge density (i.e. the difference between the spin-up and spin-down charge densities) in Figure 2a and 2b shows that the two excess electrons donated by the formation of hydroxyls mostly localize on two specific Ti atoms with d orbital character: one resides on the Ti atom of the first subsurface under an H atom; the other on the five-coordinated Ti atom. Chretien et al. had reported that such a configuration for unpaired electrons has low energy.[45] In fact, according to the ligand field arguments for octahedrally coordinated metal oxide,[46] the excess electron will occupy a Ti's 3d orbital from the $t_{2g}$ set, which will lead to the local Jahn-Teller distortion due to the electron-lattice coupling. Thus two small polarons are formed. A picture of the paramagnetic triplet state rather than the singlet closed shell state for the two excess electrons originating from a pair of hydroxyls or a bridging O removal on the $TiO_2(110)$ surface was also provided by previous calculations,[23, 25, 45] in consistent with the result of our DFT+U calcultion.

According to the density of states (DOS) of this hydroxylated surface, two peaks, which are attributed to the two polarons, present in the forbidden gap in comparison with the DOS of the pure surface, as shown in Figure 2c and 2d. One polaronic state lies about 0.50 eV and the other is about 0.77 eV below the conduction band edge. In addition, the two hydroxyl σ bonding O-H states present at about 6.23 eV below the valence band edge. Both these characteristic energy states obtained in our DFT+U calculation are in agreement with the prediction from the highly accurate hybrid exchange functional (B3LYP).[23]

**Shifts of the band edges of the hydroxylated surfaces.** Using the vacuum level as a zero energy reference, the band energies for the hydroxylated and pure surfaces are aligned. Then the differences of the valence band edge, conduction band edge, and band gap between the hydroxylated surface and the pure one are summarized in Table 1.

It is obvious that the band gap of the rutile $TiO_2(110)$ surface hardly varies with the formation of the hydroxyls on it. In fact, previous experiments performed by Wang et al. on $TiO_2$ nanowire arrays[16] and by Leshuk et al on the $TiO_2$ nanoparticles[47] did not observe the change of band gap upon the treatment by hydrogen, which seems to support our theoretical calculation. And the enhanced visible light absorption of the hydrogen-treated nanowire arrays in Wang et al's experiment[16] is be probably due to the electron transition between the polaronic states and the conduction bands.

In contrast with the almost constant band gap, the valence and conduction band edges of the surface experience the almost identical energy shift toward the vacuum level upon hydroxylation. As expected, the extent of shift increases with the hydroxyl coverage. When all outmost $O_{2c}$ atoms are bound by H atoms, the energies of both band edges increase ~1.82 eV relative to those of the pure surface. The blue shift of the conduction band edge suggests that the hydroxylated surface of this semiconductor can enhance the reducing power for producing hydrogen gas in the process of splitting water under illumination according to thermodynamics. Many experiments have reported that the hydrogen-treated $TiO_2$ and other metal oxides can exhibit the high photoactivity and solar-to-hydrogen conversion efficiency,[13-18] while the underlying mechanism is still well established. The results from our DFT+U calculations provide a reasonable explanation for the higher solar-to-hydrogen conversion efficiency of the hydrogen-treated metal oxides.

Table 1. The calculated changes in the valence band edge, the conduction band edge and the band gap ($\Delta E_V$, $\Delta E_C$ and $\Delta E_g$) of the surface with different hydroxyl coverage; positive number indicates an increase in energy with respect to the pure surface. The blue dots in squares denote the hydroxyls' locations at the surface (top view). EDM is the effective dipole moment defined in the text.

|  | 1/6 ML | 1/3 ML 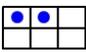 | 1/3 ML 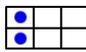 | 1/3 ML 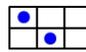 | 2/3 ML | 1 ML |
|---|---|---|---|---|---|---|
| $\Delta E_V$ (eV) | 0.51 | 0.64 | 0.77 | 0.78 | 1.17 | 1.82 |
| $\Delta E_C$ (eV) | 0.59 | 0.67 | 0.82 | 0.83 | 1.25 | 1.82 |
| $\Delta E_g$ (eV) | 0.08 | 0.03 | 0.05 | 0.05 | 0.08 | 0.0 |
| EDM (Debye/nm²) | 1.41 | 1.77 | 2.15 | 2.16 | 3.23 | 4.87 |

The hydroxyl on the TiO$_2$(110) surface is mobile.[4, 5, 48] Thus it is meaningful to study different hydroxyl sites at the same surface coverage level. As an example, the shifts of the band edges and the band gap are studied for the 1/3 monolayer (ML) hydroxyl coverage. As shown in Table 1, the results suggest the change of position of two hydroxyls does not cause much difference in the band gap as well as the band edges. In other words, the shifts of band edges due to the hydroxylation are robust, not sensitive to the hydroxyl configuration on surface.

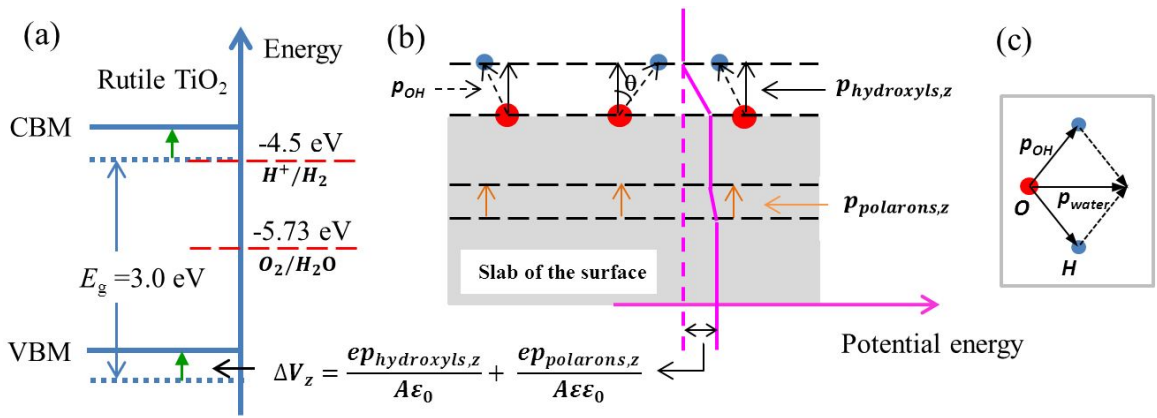

Figure 3. (a) Illustration of the shifts of the band edges of TiO$_2$. (b) The electric dipoles near the surface. These dipoles created by hydroxylation can induce the shifts of the band edges of surface TiO$_2$ in analogy to the parallel-plate capacitor. (c) The total dipole ($p_\text{water}$) and hydroxyl dipole ($p_\text{OH}$) moments of a water molecule. The angle between two hydroxyls is 104.5°, and $p_\text{water}$ is the vector sum of two $p_\text{OH}$'s.

**Effect of the electrostatic dipoles.** The blue shift of band edges and constant band gap can be attributed to the electrostatic dipoles created by the hydroxylation. The stoichiometric rutile TiO$_2$(110) surface itself is non-polar. However, for the hydroxylated surface, the total dipole moment should consist of two contributions: the intrinsic dipole of hydroxyls themselves, $p_{hydroxyls}$, and the polaronic dipoles, $p_{polarons}$, created by structural distortion and charge rearrangement. From electrostatic arguments, the z-component of the dipoles (perpendicular to the surface) can be expressed as[43]

$$p_z = p_{hydroxyls,z} + p_{polarons,z} . \tag{1}$$

The dipoles will generate a local electric field and thus the electrostatic potential will be modulated near the surface. Approximately, the variation of band edges induced by dipoles can be formulated according to the simple parallel-plate capacitor model,[49] which reads as:

$$\Delta V_z = \frac{e p_{hydroxyls,z}}{A \varepsilon_0} + \frac{e p_{polarons,z}}{A \varepsilon_0 \varepsilon} , \tag{2}$$

where A is the surface area and $\varepsilon$ is the effective dielectric constant of the surface layers. Since the chemical environment of the dipoles are different, Eq. 2 is partitioned into two terms, corresponding to the two parallel-plate capacitors – the hydroxyl one exposed in vacuum and the polaronic one embedded in the slab, as illustrated in Figure 3b. To estimate the individual dipole moment of the hydroxyl groups, $p_{hydroxyl,z}$, the dipole moment of the water molecule is used because the bond length of the hydroxyl groups on surfaces (~0.97 Å) are almost identical to that in the water molecule (0.96 Å).[50] It is well known that the total dipole moment of a water molecule, $p_{water}$, is 0.386 eÅ.[50] Thus the individual component, $p_{OH}$, can be straightforwardly estimated as 0.315 eÅ (see Figure 3c for the geometry of $p_{water}$ and $p_{OH}$). Considering the orientations of the hydroxyl groups on the surface, $p_{hydroxyls,z} = \sum p_{OH} \cdot \cos\theta_i$, where $\theta_i$ is the angle between $p_{OH}$ of a hydroxyl group and the normal of the surface. Then, for the surface with 1/6, 1/3, 2/3 and 1 ML hydroxyl coverage $p_{hydroxyls,z}$ are 0.32, 0.40 (0.49, 0.50), 0.75 and 1.14 eÅ respectively. Then the corresponding dipole moments of the polarons is 0.07, 0.07 (0.06, 0.06), 0.04 and 0.05 eÅ, obtained from the total dipole moment of the hydroxylated surface subtracting the estimated hydroxyl dipole moment. The direction of the polaron dipole moment is parallel with the hydroxyl dipole moment. The variation of band edges is in proportional to the effective dipole moment (EDM), defined by $(\varepsilon p_{hydroxyls,z} + p_{polarons,z})/\varepsilon A$. The effective dipole moments of various hydroxylated $TiO_2(110)$ surfaces are given in the last row of Table 1.

The shifts of band edges of the hydroxylated surface calculated using both Eq. 2 and aforementioned DFT+U method are plotted in Figure 4 as a function of the effective dipole moment. First, these two methods give coincident results of band edges' shifts. In this sense, it is reasonable to argue that the electric field confined in the parallel-plate capacitors, as sketched in Figure 3, results into an increase in the potential energy of the electron, and the shifts of band edges are electrostatic in origin. Second, Figure 4 also indicates the different contribution of the two components of the total dipole moment to the shifts of the band edges. While the hydroxyl groups on the surface plays a dominant role in shifts of the band edges of $TiO_2$, the component, $p_{polarons,z}/\varepsilon\varepsilon_0 A$, contributes a little and hardly varies with the hydroxyl coverage. The bridging oxygen hydroxyl coverage can effectively tune the band edges, even at a low coverage (e.g. 1/6 ML). Then the band edges of $TiO_2$ will straddle the water redox potentials well.

Finally, it should be noted that although $TiO_2$ annealed in hydrogen atmosphere may achieve a high hydroxyl coverage on the surfaces,[13] other defects may be also introduced, for example, $Ti_{5c}$-H. A $Ti_{5c}$-H defect creates the opposite dipole moment in direction relative to the $O_{2c}$-H defect, decreasing the shifts of the band edges and then the reducing power of the conduction band, which is disadvantaged for the solar-to-hydrogen conversion efficiency.

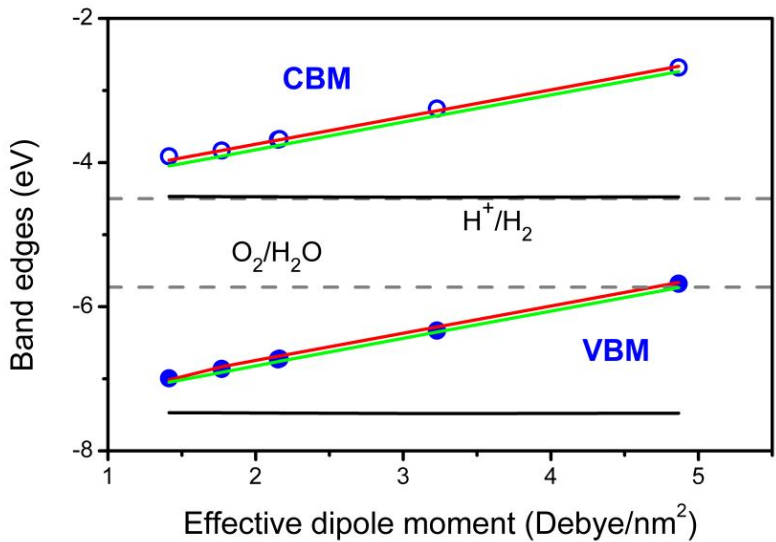

Figure 4. DFT+U energies of the conduction and valence band edges (open and solid blue symbols, respectively) of the hydroxylated $TiO_2(110)$ surfaces as a function of effective

dipole moment. The electrochemical potentials for two redox half reactions of water decomposition at pH = 7, represented by dashed lines, are shown for comparison. The solid lines represent the energy shifts of the band edges calculated from Eq. 2. Red: total contribution; Green: contributed by the hydroxyls; Black: contributed by the polaronic dipoles.

**CONCLUSIONS**

In this work, we have studied of the effect of hydroxylation of the $TiO_2$(110) surface to its band gap and band edges. While the band gap hardly varies with the extent of hydroxylation, the valence and conduction band edges experience almost identical energy shifts toward the vacuum level upon hydroxylation. The shifts of band edges are electrostatic in origin and result from dipoles created in the surface by hydroxylation. The increased conduction band edge energy means that the hydroxylated $TiO_2$ surface enhances the reducing power during water splitting under illumination.

**AUTHOR INFORMATION**

**Corresponding Author**

*E-mail: sdong@seu.edu.cn. Tel. & Fax: +86 25 52090606.

**Notes**

The authors declare no competing financial interest.

**ACKNOWLEDGMENT**

Work was supported by the NSFC (Nos. 51322206 and 11274060) and the 973 Projects of China (No. 2011CB922101).